\documentclass[superscriptaddress,amsmath,amssymb,aps,notitlepage,twocolumn,floatfix]{revtex4-2}

\usepackage{cmbright}
\DeclareFontShape{OT1}{cmss}{m}{it}{<->ssub*cmss/m/sl}{}

\DeclareFontFamily{OT1}{cmbr}{\hyphenchar\font45 }
\DeclareFontShape{OT1}{cmbr}{m}{n}{%
	<-9>cmbr8
	<9-10>cmbr9
	<10-17>cmbr10
	<17->cmbr17
}{}
\DeclareFontShape{OT1}{cmbr}{m}{sl}{%
	<-9>cmbrsl8
	<9-10>cmbrsl9
	<10-17>cmbrsl10
	<17->cmbrsl17
}{}
\DeclareFontShape{OT1}{cmbr}{m}{it}{%
	<->ssub*cmbr/m/sl
}{}
\DeclareFontShape{OT1}{cmbr}{b}{n}{%
	<->ssub*cmbr/bx/n
}{}
\DeclareFontShape{OT1}{cmbr}{bx}{n}{%
	<->cmbrbx10
}{}

\usepackage[noindentafter]{titlesec}
\usepackage{titlesec}
\titleformat{name=\section}
{\normalfont\large\bfseries\MakeUppercase}{\MakeUppercase{\thesection}}{0pt}{}
\titleformat{name=\subsection}
{\normalfont\bfseries}{\thesection}{0pt}{}
\titlespacing{\section}{0cm}{0.7cm}{0.01cm}
\titlespacing{\subsection}{0cm}{0.45cm}{0cm}

\usepackage[utf8]{inputenc}
\usepackage{listings}
\usepackage{footmisc}
\usepackage{enumerate}
\usepackage{latexsym}
\usepackage{braket}
\usepackage[version=4]{mhchem}
\usepackage{graphicx}
\usepackage[caption=false]{subfig}
\usepackage[colorlinks=True,linkcolor=red,citecolor=blue,urlcolor=blue]{hyperref}
\usepackage{blkarray}
\usepackage{array}
\usepackage[dvipsnames]{xcolor}
\usepackage[normalem]{ulem}
\usepackage{wasysym}
\usepackage{multirow}

\usepackage{pdfpages}
\makeatletter
\AtBeginDocument{\let\LS@rot\@undefined}
\makeatother

\usepackage{xspace}

\usepackage{tabularx}
\usepackage{array}   
\newcolumntype{L}{>{$}l<{$}} 
\newcolumntype{R}{>{$}r<{$}} 
\newcolumntype{C}{>{$}c<{$}} 

\usepackage{float}

\usepackage[capitalise]{cleveref}
\usepackage{placeins}

\date{\today}

\usepackage{comment}

\begin{document}
	\title{
		A j\textsubscript{eff}=1/2 Kitaev material on the triangular lattice: The case of NaRuO\textsubscript{2}
	}
	
	\author{Aleksandar Razpopov}
	\affiliation{Institut f\"ur Theoretische Physik, Goethe-Universit\"at, 60438 Frankfurt am Main, Germany}
	\author{David A. S. Kaib}
	\affiliation{Institut f\"ur Theoretische Physik, Goethe-Universit\"at, 60438 Frankfurt am Main, Germany}

	\author{Steffen Backes}
	\affiliation{Research Center for Advanced Science and Technology, University of Tokyo, Komaba, Tokyo 153-8904, Japan}
	\affiliation{Center for Emergent Matter Science, RIKEN, Wako, Saitama 351-0198, Japan}
	\affiliation{CPHT, CNRS, École polytechnique, Institut Polytechnique de Paris, 91120 Palaiseau, France}
	
	\author{Leon Balents}
	\affiliation{Kavli Institute for Theoretical Physics, University of California, Santa Barbara, California 93106, USA}
	
	\author{Stephen D. Wilson}
	\affiliation{Materials Department, University of California, Santa Barbara, California 93106-5050, USA}
	
	\author{Francesco Ferrari}
	\affiliation{Institut f\"ur Theoretische Physik, Goethe-Universit\"at, 60438 Frankfurt am Main, Germany}
	\author{Kira Riedl}
	\affiliation{Institut f\"ur Theoretische Physik, Goethe-Universit\"at, 60438 Frankfurt am Main, Germany}
	\author{Roser Valent\'i}
	\affiliation{Institut f\"ur Theoretische Physik, Goethe-Universit\"at, 60438 Frankfurt am Main, Germany}
	
	\date{\today}
	
	\begin{abstract}
		Motivated by recent reports of a quantum disordered ground state in the triangular lattice compound NaRuO$_2$, we derive a $j_{\rm eff}=1/2$ magnetic model for this system by means of first-principles calculations. The pseudospin Hamiltonian is dominated by bond-dependent off-diagonal $\Gamma$ interactions, complemented by a ferromagnetic Heisenberg exchange and a notably \emph{antiferromagnetic} Kitaev term. 
		In addition to bilinear interactions, we find a sizable four-spin ring exchange contribution with a \emph{strongly anisotropic} character, which has been so far overlooked when modeling Kitaev materials. The analysis of the magnetic model, based on the minimization of the classical energy and exact diagonalization of the quantum Hamiltonian, points toward the existence of a rather robust easy-plane ferromagnetic order, which cannot be easily destabilized by physically relevant perturbations. 
	\end{abstract}

	\maketitle
	
	\section{Introduction}
	
	The first definition of a quantum spin liquid (QSL) state dates back to the milestone paper by P.W.\ Anderson in 1973~\cite{anderson1973resonating}, in which the \textit{resonating valence-bond} wave function, a macroscopic liquid-like superposition of singlet states, was proposed as a variational guess for the ground state of the triangular lattice Heisenberg antiferromagnet~\cite{fazekas1974ontheground,balents2010spin}. Another archetypal portrait of a quantum spin liquid state is more recent and originated from the seminal work of A.\ Kitaev in 2006~\cite{kitaev2006anyons}, who set a bond-anisotropic spin model on the honeycomb lattice with an exact spin-liquid ground state represented in terms of Majorana fermions. Both these alternative descriptions of QSL states, which are associated to different microscopic mechanisms of frustration, left an indelible mark in the context of frustrated magnetism. On the one hand, the triangular lattice antiferromagnet is the prototypical example of a system with \textit{geometric frustration}, where the presence of antiferromagnetic Heisenberg couplings over odd-sided loops of sites fights the tendency towards long-range magnetic order. On the other hand, the possibility of realizing \textit{anisotropic interactions} as a consequence of the interplay between spin-orbit coupling (SOC), crystal field splitting and Hund's coupling~\cite{khaliullin2004low,khaliullin2005orbital,jackeli2009mott} has opened a whole new field of investigation centered around the Kitaev materials~\cite{winter2017models,trebst2022kitaev}. 
	Even though the original Kitaev honeycomb model has no odd-sided loops and hence no geometric frustration, it nevertheless features \textit{exchange frustration} \cite{Janssen_2019,trebst2022kitaev} due to the fact that bond-directional interactions with competing quantization axes cannot be  satisfied simultaneously. 
	
\begin{figure*}[]
		\centering
		\includegraphics[width=\linewidth]{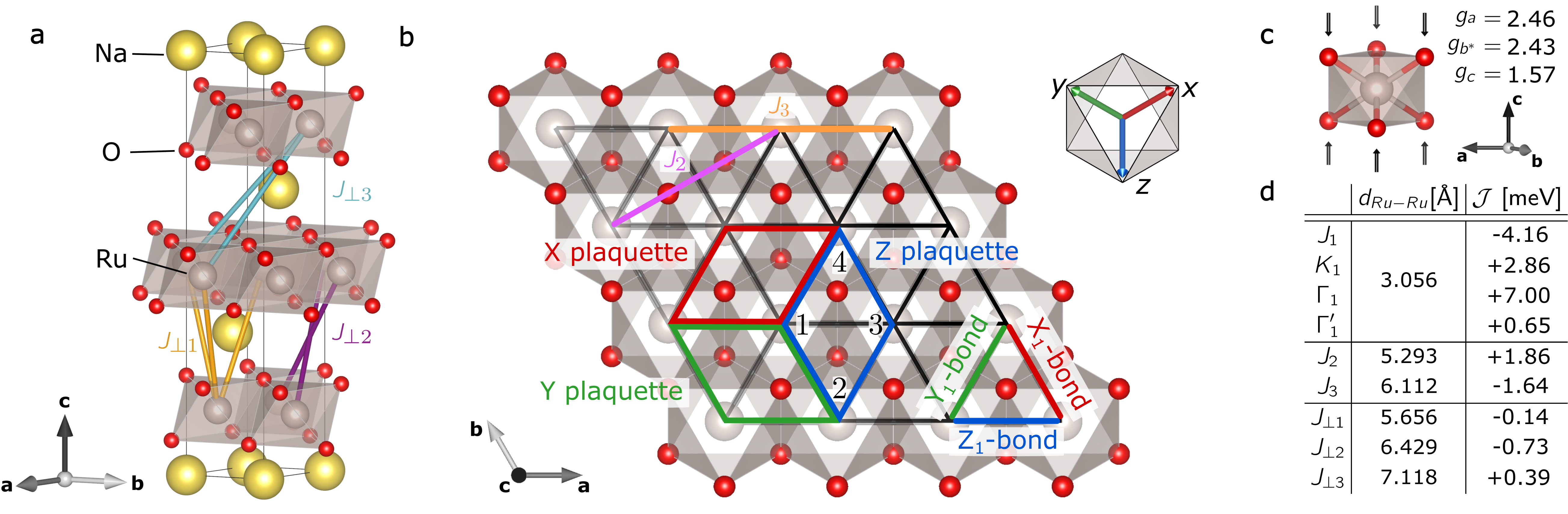}
		\caption{\textbf{NaRuO$_2$ crystal structure, illustrations and DFT results for magnetic exchange parameters.} \textbf{a} Crystal structure of NaRuO$_2$ with illustration of interlayer bonds $J_{\perp i}$. \textbf{b} Top view of the triangular Ru lattice, with illustration of intralayer bonds and four-spin ring exchange plaquettes. The cubic coordinates employed in the magnetic model are oriented approximately along Ru-O bonds, 
			as shown in the top right corner. \textbf{c} Trigonally compressed RuO$_6$ octahedron and corresponding \textit{ab-initio} $g$-tensor values, with $b^*$ defined perpendicular to the crystallographic $a$ and $c$ axes. \textbf{d} Magnetic exchange parameters for 
			nearest neighbor ($\mathcal{J}_1$) and isotropic longer-range intra- ($J_i$) and interlayer ($J_{\perp i}$) bonds. The NN couplings, as defined in \cref{eq:mag-hamiltonian}, are extracted with the projED method, while the isotropic $J$ couplings are determined by TEMA within the VASP framework. The TEMA results are scaled by a factor of $J_1^{\rm projED}/J_1^{\rm TEMA}=0.65$
			(see main text).
		}
		\label{fig:structure}
	\end{figure*}

	In this work, we investigate a recently synthesized compound, NaRuO$_2$, in which both paradigms of magnetic frustration described above come into play. The crystal structure of this material displays perfect triangular lattice planes of edge-sharing RuO$_6$ octahedra, separated by Na ions~\cite{shikano2004naruo2,ortiz2022quantum,ortiz2022defect} (illustrated in \cref{fig:structure}). 
	The same structural arrangement is found in a family
	of rare-earth chalcogenides which have been recently investigated as possible spin liquid candidates~\cite{liu2018rare}: NaYbO$_2$~\cite{bordelon2019field}, NaYbS$_2$~\cite{baenitz2018planar,sarkar2019quantum} and NaYbSe$_2$~\cite{dai2021spinon}.
	However, at variance with the latter, the space group of NaRuO$_2$ is $R\bar{3}$ and, most importantly, the rare-earth ion is replaced by ruthenium, which belongs to the $d$-block of the periodic table. 
	In analogy to the intensively studied honeycomb compound $\alpha$-RuCl$_3$~\cite{winter2017models,trebst2022kitaev}, the strong spin-orbit coupling of ruthenium, combined with the geometry of edge-sharing ligand octahedra, is expected to realize a prime example for the Jackeli-Khaliullin mechanism to form a $j_{\rm eff}=1/2$ magnet with significant Kitaev interaction~\cite{jackeli2009mott}.
	Resistivity measurements identified NaRuO$_2$ to be indeed insulating, with a small magnetization upon application of an external magnetic field, and a paramagnetic Curie temperature dependence of the magnetic susceptiblity~\cite{ortiz2022quantum}. These signatures point toward the possibility of a QSL ground state, making a microscopically motivated magnetic model for NaRuO$_2$ not only intriguing, but also necessary.

	The interplay between Heisenberg exchange and Kitaev interactions on the triangular lattice~\cite{kimchi2014kitaev,jackeli2015quantum} has been investigated in several works, revealing, for instance, the presence of crystals of $\mathbb{Z}_2$ vortices in proximity of the magnetic phase with $120^\circ$ order~\cite{rousochatzakis2016kitaev}, and possibly a spin nematic state around the antiferromagnetic Kitaev point~\cite{becker2015spin,li2015global,shinjo2016dmrg}. Furthermore, an extended spin model featuring the $\Gamma$-exchange coupling \cite{wang2021comprehensive}, which can favor the onset of a stripy magnetic phase, has been investigated in connection with the $j_{\rm eff}=1/2$ iridate compound Ba$_3$IrTi$_2$O$_9$~\cite{dey2012spin,catuneanu2015magnetic}. In this regard, a comprehensive overview of the different phases induced by bond-anisotropic (nearest-neighbor) couplings on the triangular lattice is provided by Ref.~\cite{maksimov2019anisotropic,*maksimov2019anisotropicerr}. More recently, analogous anisotropic spin Hamiltonians have been shown to capture the effective magnetic interactions of certain transition metal dihalides~\cite{stavropoulos2019microscopic,amoroso2020spontaneous,riedl2022microscopic}. 
	
	In addition to bilinear spin couplings, several magnetic materials with a triangular lattice structure, e.g.\ organic charge-transfer salts~\cite{riedl2022ingredients}, are characterized by non-negligible four-spin ring exchange interactions~\cite{thouless1965exchange}, which incorporate higher order contributions in the perturbation-theory treatment of the Hubbard model around the Mott insulating regime. While at the (semi-)classical level ring-exchange can induce the formation of spirals and non-trivial chiral orders (e.g. spin-vortex crystals)~\cite{riedl21spin}, or significantly affect the low-energy magnon spectra of collinear phases~\cite{holt14spin}, at the quantum level it is argued to potentially stabilize QSL phases~\cite{misguich1999spin,block2011exotic}; in this regard, the possible appearance of a gapless QSL with a spinon Fermi surface~\cite{motrunich2005variational}, or a Kalmeyer-Laughlin chiral state~\cite{cookmeyer2021four}, has been discussed. An additional level of complication arises in magnetic systems with strong spin-orbit interactions, namely the presence of spin-anisotropic ring-exchange interactions, which have been scarcely investigated in the past~\cite{li2022ring}.
	
	In this work we perform a thorough inspection of the magnetic properties of NaRuO$_2$, from first-principles calculations to microscopic spin models. We provide theoretical justification for the low-energy description of NaRuO$_2$ in terms of $j_{\rm eff}=1/2$ pseudospin degrees of freedom, highlighting the importance of different sources of interactions, such as intra- and inter-layer exchange couplings, and bond-anisotropic bilinear and ring-exchange interactions stemming from the strong spin-orbit coupling effects. The analysis of (classical and quantum) magnetic models indicate the existence of a robust easy-plane ferromagnetic (FM) order, which cannot be easily destabilized by perturbations around the \textit{ab initio} derived spin Hamiltonian. Based on our proposed magnetic model, we also provide theoretical inelastic neutron scattering spectra, which can be directly compared to experiment.

	\vspace{0.2cm}
	
	\section{Results}
	
	\vspace{0.2cm}{\bf Electronic properties}
	
	\vspace{0.2cm}
	
	\begin{figure}[t]
		\centering
		\includegraphics[width=1.0\linewidth]{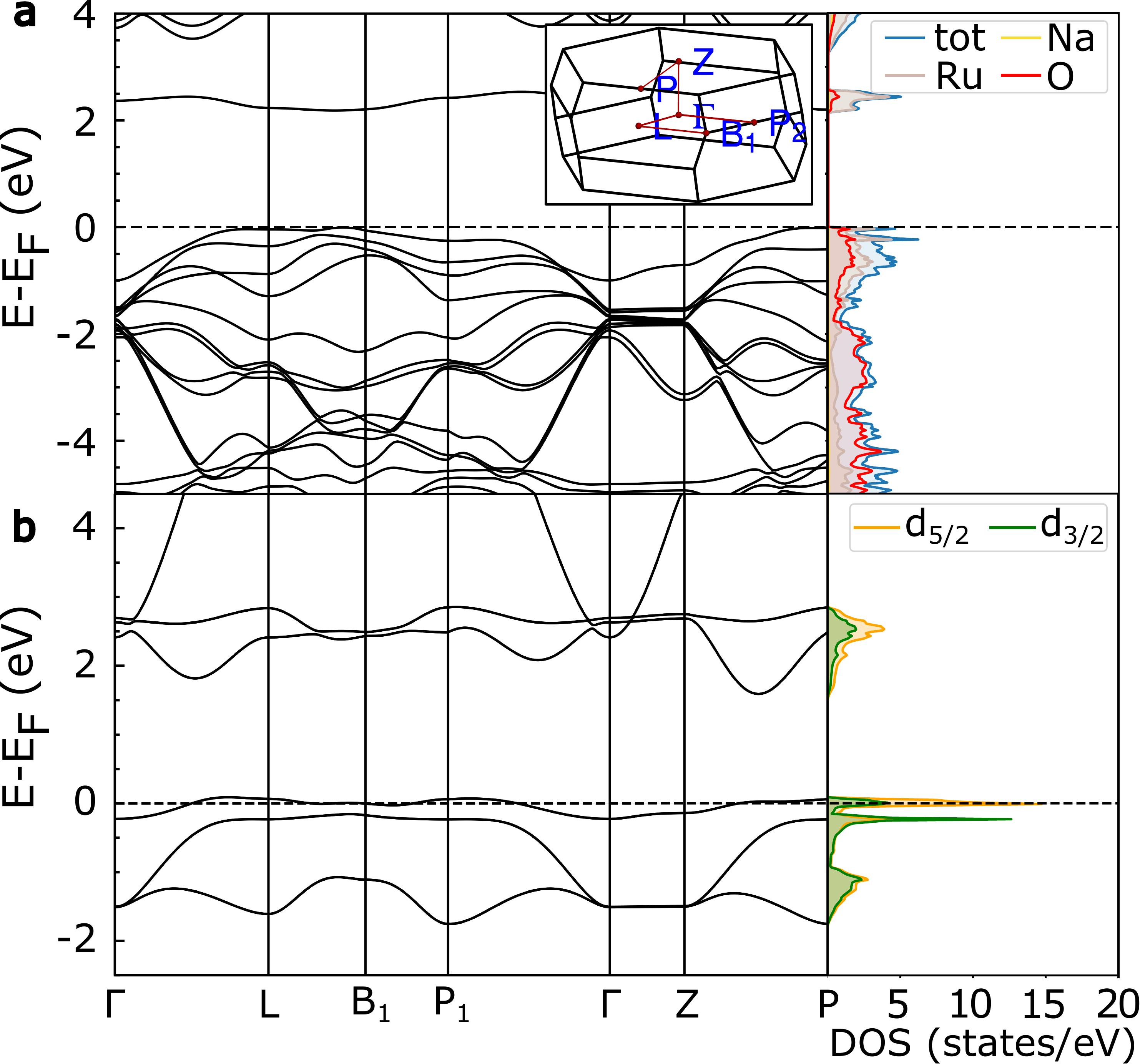}
		\caption{\textbf{Electronic band structure and density of states of NaRuO$_2$.} \textbf{a} Ferromagnetic GGA+SOC+U results, with $(U_{\rm avg},J_{\rm avg})= (3.15, 0.47)$\,eV for the Ru $4d$ electrons, and magnetization polarized in the crystallographic $a$-$b$ plane. 
			\textbf{b} Non-magnetic GGA+SOC results. The partial DOS for Ru $4d$ orbitals displays a clear splitting of $j=5/2$ and $j=3/2$ weight.}
		\label{fig:dos}
	\end{figure}

	We begin by analyzing the electronic properties of NaRuO$_2$ with the help of density functional theory (DFT) calculations, as detailed in the Methods section.
	The octahedral environment of the Ru 4$d^5$  sites leads to a crystal field splitting~\cite{winter2017models,kim2008novel,johnson2015monoclinic} with unoccupied $\rm e_g$-states and occupied t$_{\text{2g}}$-states. The latter further split into $j_{\text{eff}}$ = 3/2 and $j_{\text{eff}}$ = 1/2 levels in the limit of strong spin-orbit coupling. 
	
	To estimate the Hubbard repulsion and Hund's coupling of NaRuO$_2$, we employ constrained random phase approximation (cRPA) calculations (see Methods section).
	In the non-relativistic band structure we observe a crossing between the Ru $e_g$ band and a band with dominant Na $3s$ character close to the $\Gamma$ point. This poses the question about which bands should be included in the Wannierization procedure prior to cRPA. One option is considering only the five $4d$ Ru bands, which can be expected to lead to an artificially enhanced $e_g$ screening. The corresponding cRPA result is $(U_{\rm avg},J_{\rm avg})_{4d}=(3.114,0.4736)\,$eV. Alternatively, the crossing band may be included in a cRPA calculation based on a six-band model. This choice leads to an artificially suppressed screening, with $(U_{\rm avg},J_{\rm avg})_{4d+3s}=(3.1865,0.4756)\,$eV. 
	Since it turns out that both approaches lead to similar results, we choose to work with the average of them: $(U_{\rm avg},J_{\rm avg})=(3.15,0.47)\,$eV.
	
	We employ these values as correlation corrections on the Ru $4d$ electrons in a relativistic band structure calculation 
	(GGA+SOC+U, see Methods section) with ferromagnetically aligned magnetic moments. The resulting band structure and partial density of states (DOS) are shown in \cref{fig:dos}a. We choose the spin magnetic moments to be polarized in the crystallographic $a$-$b$ plane, which is the most energetically favorable orientation.
	Within this setting we obtain an insulator with a charge gap of 2.1\,eV which is in agreement with recent resistivity experiments~\cite{ortiz2022quantum}.  
	We find that magnetism and Coulomb interaction are necessary to open a realistic charge gap in NaRuO$_2$, which is further enlarged by SOC. 
	
	The partial DOS resolves the dominance of ruthenium weight around the Fermi level, such that we can proceed with a low-energy modelling of this compound based on ruthenium bands.
	
	The edge-sharing octahedral structure of NaRuO$_2$ hints towards the possibility of a $j_{\rm eff}=1/2$ description of the low-energy magnetic properties, in analogy with the intensively studied $\alpha$-RuCl$_3$~\cite{winter2017models}.
	To check the validity of the relativistic $j_{\rm eff}$ picture for NaRuO$_2$, we calculate also a non-magnetic band structure and a partial DOS resolved with respect to the general relativistic basis for $d$ orbitals, $j=\{5/2,3/2\}$ (shown in \cref{fig:dos}b). As detailed in Supplementary Note~1, the $j=3/2$ states can be directly expressed as a sum of $j_{\rm eff}=3/2$ and $e_g$ states. The information about the $j_{\rm eff}=1/2$ states on the other hand is contained in the $j=5/2$ states. The small contribution of $j=3/2$ states around the Fermi energy in \cref{fig:dos}b is thus an indication that a description in terms of pseudospin $j_{\rm eff}=1/2$ states is reasonable.

	\vspace{0.2cm}
	
	\vspace{0.2cm}{\bf Magnetic Model}
	
	\vspace{0.2cm}
	
	As an appropriate magnetic model for NaRuO$_2$ we consider a $j_{\rm eff}=1/2$ pseudospin Hamiltonian. 
	To relate the pseudospin $\mathbf S$ of the magnetic Hamiltonian to the magnetic moment, $\mathbf M = \mu_\text{B} \mathbb G \cdot \mathbf S$, we calculate the gyromagnetic $g$-tensor from first principles (see Methods section). We find it to be approximately diagonal with $(g_a,g_{b^\ast},g_c)=(2.46,2.43,1.57)$ in crystallographic coordinates (with $b^\ast$ perpendicular to $a$ and $c$).
	With respect to the triangular plane, the in-plane components ($g_a,g_{b^\ast}$) are larger than the out-of-plane one ($g_c$), as a direct consequence of the trigonal compression of the RuO$_6$ octahedra along the crystallographic $c$ axis (see Fig.~\ref{fig:structure}c)~\cite{chaloupka2016magnetic}. 
	
	For the magnetic interactions between the pseudospins we consider a Hamiltonian consisting of a bilinear exchange term $\mathcal{H}_{2}$ and a four-spin ring exchange term $\mathcal{H}_{4}$. We express this model in the conventionally used cubic coordinates for Kitaev materials, which consist of orthogonalized axes oriented approximately along the Ru-O bonds, as illustrated in the top right corner of \cref{fig:structure}b. We denote the three components of the pseudospin at site $i$ as $S^\mu_i$, with $\mu=\{x,y,z\}$.  In this framework, the $[111]$ pseudospin direction is parallel to the crystallographic $c$ axis. For completion, in Supplementary Note~2 we translate our model to an alternative reference frame with crystallographic coordinates~\cite{maksimov2019anisotropic,*maksimov2019anisotropicerr}.
	
	The bilinear contribution to the magnetic Hamiltonian $\mathcal{H}_{2}=\sum_{i<j} \sum_{\mu\nu} \mathbb{J}_{i,j}^{\mu\nu} {S}_i^\mu   {S}_j^\nu$ contains, especially for nearest neighbors, anisotropic bond-dependent terms. Considering the symmetry constraints of the $R\bar{3}$ space group, the bilinear exchange tensor on a Z$_1$-bond (as defined in \cref{fig:structure}b) follows the form
	\begin{align}
		\mathbb{J}_\mathrm{Z_1-bond} = \begin{pmatrix}
			J_1 & \Gamma_1 & \Gamma^\prime_1 \\
			\Gamma_1 & J & \Gamma^\prime_1 \\
			\Gamma^\prime_1 & \Gamma^\prime_1 & J_1 + K_1
		\end{pmatrix}\label{eq:mag-hamiltonian}.
	\end{align}
	Here, $J_1$ is the isotropic Heisenberg exchange, $K_1$ the Kitaev coupling, and $\Gamma_1$ and $\Gamma^\prime_1$ the off-diagonal symmetric exchange parameters. 
	The bilinear interactions on X$_1$- and Y$_1$-bonds are then related to this expression by $C_3$ spin rotations around the [111] axis, amounting to cyclic permutation of $(x,y,z)$ spin components. 
	
	We performed DFT calculations to obtain the magnetic exchange parameters as described in the Methods section.
	In \cref{fig:structure}d we show the dominant magnetic couplings extracted
	for nearest-neighbor bonds with the projED method~\cite{riedl2019abinitio} and the isotropic longer-range exchange from total energy mapping analysis (TEMA).
	Both methods predict similarly strong intralayer ferromagnetic Heisenberg $J_1$ coupling, when we consider a scaling factor $J_1^{\rm projED}/J_1^{\rm TEMA} \approx 0.65$.
	The shortest further-neighbor intralayer couplings are found to be non-negligible and of similar magnitude, with $J_2$ being anti- and $J_3$ ferromagnetic.
	The interlayer Heisenberg couplings $J_{\perp 1}$, $J_{\perp 2}$ and $J_{\perp 3}$ (shown in Fig.~\ref{fig:structure}a) are one magnitude smaller than the intralayer ones, with $J_{\perp 1}$ and $J_{\perp 2}$ being ferromagnetic and $J_{\perp 3}$ antiferromagnetic.
	
	Within the projED method we obtain the bond-dependent anisotropic couplings $K_1$, $\Gamma_1$ and $\Gamma^\prime_1$ at nearest-neighbors, where the cRPA values $(U_{\rm avg},J_{\rm avg})=(3.15,0.47)$\,eV are employed. 
	Compared to previously estimated magnetic parameters for other Ru $4d$ systems~\cite{winter2017breakdown,winter2016challenges,wolf2022strongly,kaib2022electronic}, it is interesting to note that we have strongly \textit{antiferromagnetic} Kitaev $K_1$ term and a dominant positive $\Gamma_1$ as the largest coupling.
	These are contributions which, to the best of our knowledge, have not been observed in a real material with effective spin~$1/2$ so far.
	
	The microscopic origin of the unusual antiferromagnetic sign of the Kitaev interaction encountered here can be understood as follows: From the perspective of second-order perturbation theory in a perfect octahedral environment (considering only the occupied $t_{2g}$ orbitals), the Kitaev interaction scales as ${K_1 \propto (t_1-t_3)^2-3\,t_2^2}$~\cite{rau2014generic}. On a Z-bond, the hopping parameters are defined as the ligand-assisted hopping $t_2=t_{(xz;yz)}$, as well as $t_1=t_{(xz;xz)}=t_{(yz;yz)}$ and $t_3=t_{(xy;xy)}$, which stem predominantly from direct $d$ orbital overlap. 
	For the prime example of the honeycomb Kitaev material $\alpha$-RuCl$_3$, the indirect hopping $t_2$ is dominant, somewhat close to the $t_2$-only model in the Jackeli-Khaliullin mechanism, where the Kitaev interaction is ferromagnetic ($K_1\propto -3t_2^2$)~\cite{jackeli2009mott}. In comparison, the direct hoppings $t_3$ and $t_1$ gain importance in NaRuO$_2$, where the nearest neighbor Ru-Ru bond length is significantly smaller than in $\alpha$-RuCl$_3$. 
	As dictated by the geometry, $t_3$ is negative and larger in magnitude than the positive $t_1$. The resulting \emph{antiferromagnetic} Kitaev interaction in NaRuO$_2$ can hence be directly related to the shorter nearest-neighbor bond length of this triangular compound. 
	The perturbation theory perspective also offers an explanation for the dominance of the off-diagonal symmetric exchange $\Gamma_1$, which scales approximately as $\Gamma_1 \propto t_2\,(t_1-t_3)$~\cite{rau2014generic}. While the magnitude of $K_1$ reduces with the competition between indirect and direct contributions, $\Gamma_1$ increases proportional to the magnitude of the hoppings, leading to a magnetic model dominated by the off-diagonal symmetric  exchange for NaRuO$_2$. 
	
	Despite the antiferromagnetic sign of the Kitaev interaction, the bilinear exchange Hamiltonian $\mathcal H_2$ features a ferromagnetically ordered ground state, due to significant $J_1<0$ and $\Gamma_1>0$ interactions, as discussed in more detail below. 
	To seek out possible additional interactions that might destabilize the ferromagnetic ground state, we consider the effects of higher-order ring exchange interactions. Compared to honeycomb Kitaev materials, NaRuO$_2$ could be predestined for such interactions, due to the marginally insulating Mott state reported 
	in experiments~\cite{ortiz2022quantum} and to its triangular lattice structure, where the shortest closed loops consist of four (instead of six) sites. 
	Four-spin ring exchange without spin-orbit coupling effects has been discussed plentifully in the literature \cite{thouless1965exchange,holt14spin,misguich1999spin,block2011exotic,cookmeyer2021four} and takes the form
	\begin{equation}
		\begin{aligned}
			\label{eq:isoring}
			\mathcal H_4^{\rm iso} =  \frac{K^{\rm iso}}{S^2}  \sum_{\langle ijkl \rangle}
			&(\mathbf{S}_i \cdot \mathbf{S}_j) (\mathbf{S}_k \cdot \mathbf{S}_l)+ (\mathbf{S}_i \cdot \mathbf{S}_l) (\mathbf{S}_j \cdot \mathbf{S}_k) 
			\\ & - (\mathbf{S}_i \cdot \mathbf{S}_k) (\mathbf{S}_j \cdot \mathbf{S}_l),
		\end{aligned}
	\end{equation}
	where the summation $\langle ijkl\rangle$ goes over plaquettes with sites $i$ and $k$ lying across a diagonal~\cite{thouless1965exchange}, $K^{\rm iso}$ is the coupling constant, and the superscript ``iso'' denotes that this is the conventional isotropic (i.e., SU(2)-symmetric) ring exchange. 
	
	However, for NaRuO$_2$ there is no reason why the ring exchange between the pseudospin $j_{\rm eff}=\frac12$ moments should follow the form of $\mathcal H_4^\mathrm{iso}$, since spin-orbit coupling is expected to induce anisotropic four-site terms in the Hamiltonian. 
	In the most general form, anisotropic four-spin exchange may be expressed as
	\begin{align}
		\mathcal{H}_{4}^\mathrm{tot} = \frac{1}{S^2} \sum_{\langle ijkl \rangle} \sum_{\mu \nu  \rho \eta} {\mathbb K}_{ijkl}^{\mu \nu \rho \eta } \, (S_i^\mu S_j^\nu S_k^\rho S_l^\eta),
		\label{eq:ring-exchange}
	\end{align}
	where the tensor $\mathbb K$ contains the coupling constants.  
	The presence of inversion symmetry with respect to the center of each plaquette, together with a $C_2$ rotation axis parallel to the shortest diagonal, reduces the 81 entries of $\mathbb K$ for one plaquette to 24 independent parameters. 
	Furthermore, analogous to the X-, Y-, and Z-bonds of bilinear exchange, it is convenient to define X-, Y-, and Z-plaquettes, as shown in \cref{fig:structure}b. 
	The three plaquettes are related by $C_3$ rotations around the out-of-plane axis
	and hence the tensor of one plaquette type fully encodes $\mathcal H_4^\mathrm{tot}$. 
	
	Note that in contrast to conventional ring exchange, care has to be taken for the order of the site-numbering within a plaquette. For example, for a single plaquette (with the site labeling illustrated in \cref{fig:structure}b), swapping two sites across a diagonal is not a symmetry of the ring-exchange tensor, i.e.\ 
	{$\mathbb{ K}_{1234}^{\mu \nu \rho \eta } \neq \mathbb{ K}_{1432}^{\mu \nu \rho \eta}$}, even in presence of the aforementioned symmetries.

	To compute the ring-exchange tensor $\mathbb{K}$ from first principles, we employ the projED method, which has been used previously to determine ring-exchange couplings for organic triangular lattice compounds~\cite{riedl19critical,riedl21spin}.
	Results on a Z-plaquette are given in \cref{tab:fourspin}, and details of the calculation are outlined in the Methods section.  
	We do not attempt to create a minimal model here and show agnostically the full \textit{ab-initio} result. Overall, the four-site ring exchange contribution in NaRuO$_2$ does not seem to be obviously negligible, with {a} strength of roughly 5-10\% of the nearest-neighbor bilinear exchange parameters. 
	As anticipated, the shown results deviate substantially from the conventional isotropic ring exchange $\mathcal H_4^\mathrm{iso}$. This is not surprising because of the strong spin-orbit coupling in NaRuO$_2$. For instance, among the diagonal components characterizing the Z-plaquette, the {${\mathbb K}_{1234}^{zzzz}$} term strongly differs from {${\mathbb K}_{1234}^{xxxx}={\mathbb K}_{1234}^{yyyy}$}. 
	
	To quantify the degree of anisotropy of the total ring exchange Hamiltonian
	encoded in \cref{tab:fourspin}, we express it 
	as a sum consisting of the conventional isotropic ring exchange from \cref{eq:isoring} and a purely anisotropic contribution: $\mathbb K^{\rm tot} = \mathbb K^{\rm iso} + \mathbb K^{\rm ani}$. The choice of this splitting is not unique, but we choose the coupling constant $K^{\rm iso}$ {in $\left(\mathbb K^{\rm iso}\right)^{\mu\nu\rho\eta}_{1234}=K^\mathrm{iso}\left( \delta_{\mu\nu}\delta_{\rho\eta} + \delta_{\mu\eta}\delta_{\nu\rho} - \delta_{\mu\rho}\delta_{\nu\eta} \right)$} (cf.\ Eq.~\eqref{eq:isoring}) such that the tensor-norm of the anisotropic part, 
	$
	\|\mathbb K^{\rm ani}\| = \| \mathbb K^{\rm tot}-\mathbb K^{\rm iso} \|$, is minimized. Here, the tensor 1-norm is used ($\|\mathbb K\| = \sum_{\mu\nu\rho\eta} |\mathbb K_{1234}^{\mu\nu\rho\eta}|$). 
	This choice is motivated by an analogy to the case of the bilinear Hamiltonian, where the same procedure splits the bilinear exchange tensor $\mathbb J_{ij}^{\mu\nu}$ into an (isotropic) Heisenberg exchange part and an anisotropic part, arriving at the same definition of Heisenberg-$J$ as in Eq.~\eqref{eq:mag-hamiltonian}. Dissecting the ring exchange interaction in this way leads to $K^{\mathrm{iso}}=-0.06\,\text{meV}$ and $\|{\mathbb K^{\rm ani}}\|\,/\,\|\mathbb K^\mathrm{iso} \|= 5.6$, which shows that ring exchange in NaRuO$_2$ is dominated by the anisotropic contributions. 
	
	\begin{table}[]
		\begin{tabular}{c|c}
			$\mathbb{K}_{1234}^{\mu \nu \rho \eta}$\,[meV] & $S_1^\mu S_2^\nu S_3^\rho S_4^\eta$ \\
			\hline
			\hline
			0.1829 & ( xxzx + yyzy + zxxx + zyyy ) \\ 
			0.1828 & ( xxzy + yxzy + zyxx + zyyx ) \\ 
			0.1819 & ( zxzy + zyzx ) \\ 
			0.1810 & ( xzyz + yzxz ) \\ 
			-0.1740 & ( zzzz ) \\ 
			-0.1559 & ( xyzz + yzzx + zxyz + zzxy ) \\ 
			-0.1457 & ( xzzx + yyzz + zxxz + zzyy ) \\ 
			-0.1398 & ( xxxx + yyyy ) \\ 
			-0.1284 & ( xxxz + xzxx + yyyz + yzyy ) \\ 
			0.1122 & ( zxzz + zyzz + zzzx + zzzy ) \\ 
			-0.0969 & ( xyxy + yxyx ) \\ 
			-0.0829 & ( xyzy + yxzx + zxyx + zyxy ) \\ 
			0.0602 & ( xyzx + yyzx + zxxy + zxyy ) \\ 
			-0.0596 & ( xyxz + xzxy + yxyz + yzyx ) \\ 
			-0.0581 & ( xzzz + yzzz + zzxz + zzyz ) \\ 
			-0.0580 & ( xxzz + yzzy + zyyz + zzxx ) \\ 
			-0.0570 & ( xzyx + xzyy + yxxz + yyxz ) \\ 
			-0.0542 & ( xxxy + xyxx + yxyy + yyyx ) \\ 
			0.0354 & ( xxyz + xyyz + yzxx + yzxy ) \\ 
			-0.0352 & ( xzzy + yxzz + zyxz + zzyx ) \\ 
			-0.0272 & ( xxyy + xyyx + yxxy + yyxx ) \\ 
			0.0252 & ( xzxz + yzyz ) \\ 
			0.0145 & ( xxyx + xyyy + yxxx + yyxy ) \\ 
			0.0069 & ( zxzx + zyzy ) 
		\end{tabular}
		\caption{
			\textbf{Four-spin ring exchange couplings for NaRuO$_2$.} The corresponding ring-exchange expression is defined in \cref{eq:ring-exchange} and the parameters in this table refer to a Z-plaquette, with the site labeling  $1,2,3,4$ illustrated in \cref{fig:structure}b. The ring-exchange parameters are calculated with projED starting from a three-orbital Hubbard model. 
			For simplicity, we abbreviate $S_1^\mu S_2^\nu S_3^\rho S_4^\eta$ with ``$\mu \nu \rho \eta$''. Parameters for X- and Y-plaquettes follow by 
			$C_3$ rotations around the out-of-plane axis.
		}
		\label{tab:fourspin}
	\end{table}
	
	\vspace{0.2cm}{\bf Properties of the Magnetic Model}
	\vspace{0.2cm}
	
	We investigate the ground state of the magnetic model given in \cref{fig:structure}d by applying two different methods.
	We consider the classical ground state via an iterative minimization method of the energy~\cite{walker1980computer,sklan2013nonplanar} and then we include quantum fluctuations by tackling the Hamiltonian with exact diagonalization (ED) on finite clusters with up to 27 sites (see Methods section and Supplementary Note~3). 
	
	First we consider the classical ground state of the ${\cal H}_2$ Hamiltonian restricted to the triangular lattice plane.
	The omission of inter-layer couplings is justified by their small estimated magnitude compared to intra-layer couplings (cf. \cref{fig:structure}d). The minimum of the classical energy is provided by a ferromagnetic spin arrangement, with spins lying in the triangular lattice plane. The ferromagnetic nature of the ground state turns out to be stable upon different perturbations of the Hamiltonian around the \textit{ab initio} model (as discussed in Supplementary Note~4), and upon inclusion of anisotropic ring exchange or out-of-plane interactions. For what concerns the latter, the classical energy minimum yields a ferromagnetic ground state, with spins being parallel to each other both within and between layers, consistent with total energy calculations within DFT. However, the configuration with ferromagnetically stacked FM layers is lower in energy than the one with antiferromagnetically stacked FM layers only by $\sim$0.1 meV/Ru. ED calculations, performed on various two-dimensional clusters with different shapes and number of sites, confirm the stability of the ferromagnetic ground state when quantum effects come into play. The addition of the ring-exchange interaction $\mathcal H_4^\mathrm{tot}$ does not destabilize the ferromagnetic order, neither for classical nor for quantum spins, but leads to a small tilt of the ordered moment out of the triangular lattice plane (by less than $1^\circ$ in our model).
	
\begin{figure}
		\centering
		\includegraphics[width=\linewidth]{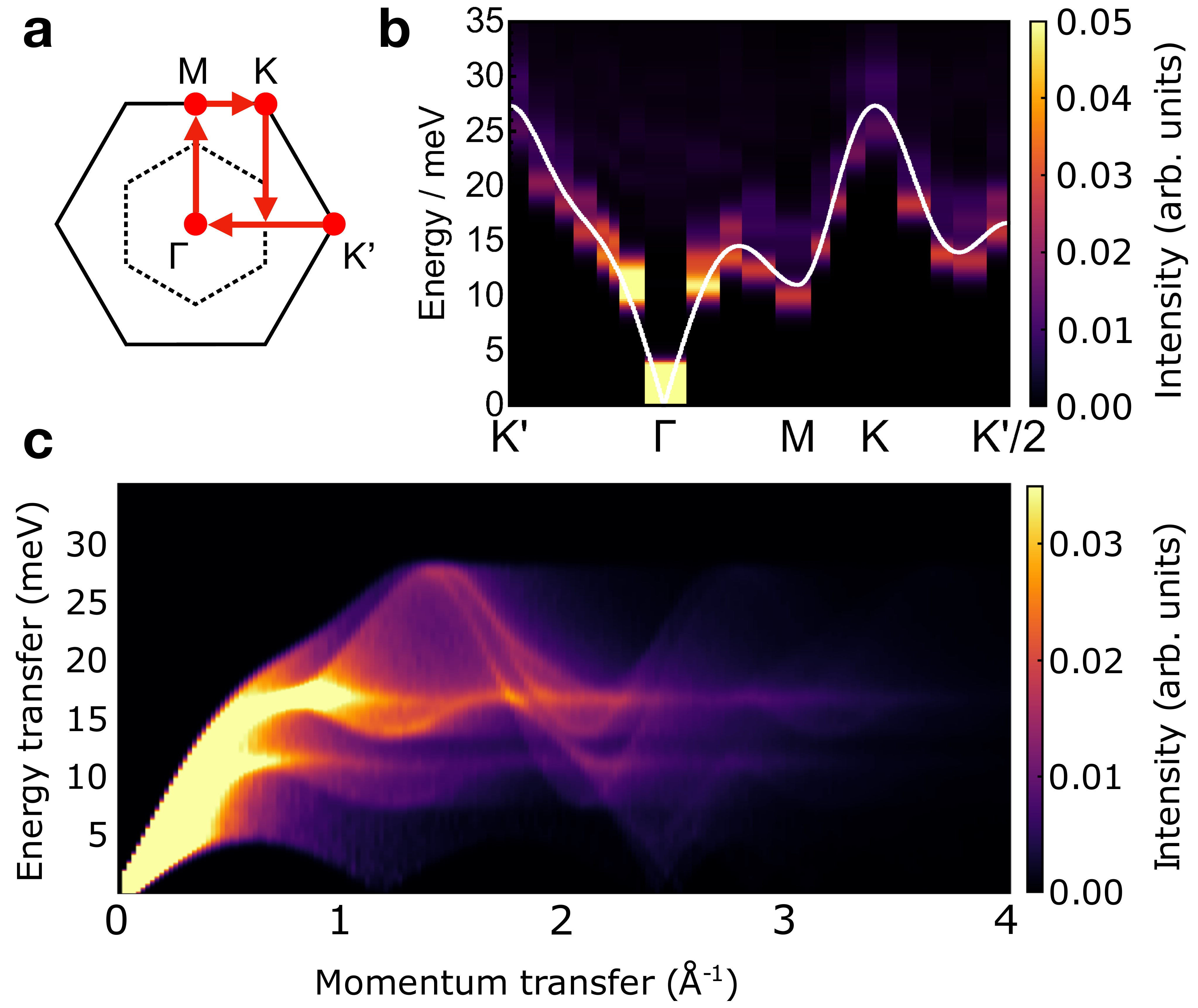}
		\caption{
			\textbf{Computed neutron scattering intensity within the magnetic model.} \textbf{a} Momentum path in the Brillouin zone of the triangular lattice, \textbf{b} INS intensity for the  Hamiltonian $\mathcal H_2$ without inter-plane interactions, along high-symmetry lines in momentum space. Color plot intensity from ED (with a broadening of 1\,meV), combined from different clusters from 18 to 27 sites. The overlayed band dispersion is the single-magnon energy from LSWT. As the ED ground state is a superposition of different degenerate magnetic domains, the compared LSWT energy is plotted as the average of three calculations, which correspond to expansions around in-plane ferromagnetic order, with magnetic moments perpendicular to X$_1$-, Y$_1$- or Z$_1$-bonds. 
			\textbf{c} Calculated powder-averaged INS spectrum, from the model in \cref{fig:structure}d, including inter-plane exchange couplings. Note that (different) arbitrary units are used for the intensities in \textbf{b} and \textbf{c}, with cut-offs due to large intensity at $\mathbf q \approx 0$. }
		\label{fig:INS}
	\end{figure}

	After having established the ferromagnetic character of the ground state, we move on to compute excitations, namely the inelastic neutron scattering (INS) intensity predicted by the magnetic model.
	We employ linear spin-wave theory (LSWT), complemented by ED to investigate effects beyond LSWT. The results are summarized in Fig.~\ref{fig:INS}, where the \textit{ab initio} magnetic form factor for Ru$^{3+}$~\cite{do2017majorana} is taken into account for the calculation of spectral intensities, such that the magnetic spectra can be directly compared to neutron scattering experiments.
	
	As previously mentioned, the magnetic moments of the ferromagnetic ground state of the $\mathcal{H}_2$ Hamiltonian lie within the triangular lattice plane, without picking any preferred direction on the classical level. However, this continuous symmetry is accidental and is lifted by quantum fluctuations, which select the configurations in which the moments are perpendicular to one of the nearest-neighbor bonds~\cite{maksimov2019anisotropic,*maksimov2019anisotropicerr}. At the \textit{linear} spin wave theory level we then expect the appearance of a gapless pseudo-Goldstone mode, which becomes gapped when quantum effects beyond the lowest-order are considered~\cite{rau2018pseudo}. The ED results, compared to the LSWT prediction in Fig.~\ref{fig:INS}b, confirm this picture, and find a gap at $\mathbf q=0$ of the order of 1~to 2~meV.
	
	A further effect beyond LSWT that one might expect here is the appearance of strong scattering continua even in magnetically ordered phases. Such continua were observed in the honeycomb Kitaev material $\alpha$-RuCl\textsubscript{3}\ \cite{banerjee2017neutron}, where they have been traced back in a spin-wave description to originate from significant anharmonic effects due to $\Gamma_1$ exchange \cite{winter2017breakdown}. 
	However, despite the dominant off-diagonal $\Gamma_1$ exchange in the case of NaRuO$_2$ (cf.\ \cref{fig:structure}d), no substantial scattering continuum is found here and the ED spectrum qualitatively follows the sharp modes of LSWT, as shown in a comparison in \cref{fig:INS}b. This can be understood as a consequence of the fact that in the present ferromagnetic state, the pseudo-Goldstone modes remain at the ordering-wave vector $\mathbf Q=0$ ($\Gamma$-point), such that a potential decay of single-magnons into a two-magnon continuum via $\Gamma_1$-exchange is kinematically not allowed.
	We note that the inclusion of ring exchange increases the magnon energies by $\sim 2 $\,meV, but does not qualitatively change the main features of the spectrum.  
	
	We also compute the powder-averaged INS spectrum that might be relevant for direct comparison of the predicted FM state to experiments. Here, the effect of inter-plane couplings are included, in order to obtain meaningful integration over out-of-plane momenta. The results from LSWT are shown in \cref{fig:INS}c and feature the gapless pseudo-Goldstone mode at smallest momenta, and a less intense gapless mode around 1.2~$\AA^{-1}$ arising from the inter-layer ferromagnetic stacking.

	\vspace{0.2cm}{\bf Conclusions and Outlook}
	\vspace{0.2cm}
	
	In this work we investigated the magnetic properties of NaRuO$_2$, a layered system of corner-sharing RuO$_6$ octahedra, which constitutes 
	a prime example for the realization of anisotropic spin couplings, such as the Kitaev interaction, on a triangular lattice structure.
	By combining two complementary first-principle methods, TEMA and projED, we derived a $j_{\rm eff}=1/2$ pseudospin Hamiltonian for NaRuO$_2$, which displays a sizable \emph{antiferromagnetic} Kitaev coupling. This is a direct consequence of the comparatively smaller nearest-neighbor Ru - Ru bond length in NaRuO$_2$, leading to a dominance of direct hopping mechanisms in contrast to other spin-$1/2$ Kitaev materials to date.
	The strongest interactions of the model are however a symmetric $\Gamma_1$ exchange and a ferromagnetic $J_1$ Heisenberg term. The spin Hamiltonian with bilinear interactions possesses a rather robust ferromagnetic order, oriented parallel to the triangular lattice plane formed by ruthenium ions, also when longer-range intra- and inter-plane exchange interactions are taken into account. 
	
	The parameters of the magnetic Hamiltonian, as obtained by first-principles calculations with the pristine structure of NaRuO$_2$, place this material deep inside an extended ferromagnetic phase, which cannot be easily destabilized by perturbing the Hamiltonian around the \textit{ab initio} point. Since the experimentally available results do not show the conventional features of long-range ferromagnetic order and suggest the possibility of a marginally insulating Mott state in NaRuO$_2$~\cite{ortiz2022quantum}, we explored the effects of higher-order spin interactions, computing four-spin ring exchange couplings from the projED method. However, although the latter turn out to be of non-negligible size, they seem insufficiently strong to melt the ferromagnetic order. Nevertheless, the nature of the ring exchange interaction is strongly anisotropic and its consequences warrant further investigation, also in the context of other Kitaev materials. 
	
	Furthermore, employing LSWT calculations and exact diagonalization on finite clusters, we computed the inelastic neutron scattering spectra for NaRuO$_2$. Comparison of the powder-average neutron scattering intensity with experimental observations shows a similar weight distribution which may signal the presence of underlying ferromagnetism in the system, although no long-range magnetic
	order was observed in experiments~\cite{ortiz2022quantum,ortiz2022defect}. This raises the question of the role of disorder in the material, which may be addressed in future investigations.

	\vspace{0.2cm}{\bf Note added}
	\vspace{0.2cm}
	
	During the process of completion of this manuscript, we became aware of the preprint~\cite{bhattacharyya2022antiferro}, which provides a nearest-neighbor spin Hamiltonian for NaRuO$_2$ from quantum chemistry methods. While the signs of the nearest-neighbor couplings match with the ones of our model, the $\Gamma_1$ term of Ref.~\cite{bhattacharyya2022antiferro} is smaller than the $J_1$ exchange strength, contrary to our results. The authors explore the possibility of destabilizing the ferromagnetic order by an antiferromagnetic third-neighbors exchange, which is in contrast with our first-principle prediction of a ferromagnetic $J_3$ coupling.
	
	\section{Methods}
	
	\vspace{0.2cm}{\bf First principles calculations}
	
	\vspace{0.2cm}
	
	All first principles calculations employ the crystal structure published in Ref.~\cite{ortiz2022quantum}.
	For the calculation of electronic properties  we use the full potential local orbital (FPLO)~\cite{fplo} package 18.00-57 and the Generalized Gradient Approximation (GGA)~\cite{GGA} as the exchange-correlation functional.
	The correlation for the strongly localized Ru 4d electrons are corrected via the GGA+U approximation using the ``atomic limit'' implementation~\cite{FPLO_AMF}.
	All calculations are carried out on a 12$\times$12$\times$12 k-grid in the primitive unit cell. Relativistic calculations are performed within the GGA+SOC+U functional.
	The results have been cross-checked with the linearized augmented plane-wave basis set as implemented in Wien2k~\cite{wien2k} version 19.1,
	with Ru $4d$ correlation correction included via the SIC method~\cite{SIC1,liechtenstein1995density} with effective Couloumb repulsion \text{$U_{\rm eff} = 2$\,eV}.
	
	We also compute the gyromagnetic $g$-tensor from first principles.
	For this calculation, we consider a [RuO$_6$]$^{9-}$ molecule within the quantum chemistry ORCA 5.03 package~\cite{neese2012orca,*neese2022orca,*neese2005efficient} with the functional TPSSh, basis set def2-TZVP and complete active space self-consistent field method for the $d$ orbitals CASSCF(5,5).
	A conductor-like polarizable continuum model (C-PCM)~\cite{barone1998cpcm} is employed with a Gaussian charge scheme, a van der Waals-type cavity and an infinite dielectric constant. 
	
	\vspace{0.2cm}{\bf Constrained random phase approximation (cRPA)}
	
	\vspace{0.2cm}
	
	Based on the electronic structure obtained with the Wien2k package v21.1~\cite{wien2k}  we estimate the electronic two-particle interaction terms in NaRuO$_2$ with the constrained random-phase approximation (cRPA)~\cite{aryasetiawan2004frequency,aryasetiawan2006calculations}, as implemented in the FHI-gap code~\cite{fhigap}. The integration of the Brillouin zone is done on an $8\times 8 \times 8$ grid. The static low-energy limit of the partially screened interaction is projected onto the relevant orbitals, where screening processes in the same window are excluded. The spherical symmetric expressions for $d$ electrons in the atomic limit are based on Slater integrals $F_k$~\cite{liechtenstein1995density} as follows:
	\begin{align}
		U_{\rm avg} &= \frac{1}{(2l+1)^2}\sum_{\alpha \beta} U_{\alpha \beta}=F_0 \\
		J_{\rm avg} &=\frac{7}{5} \frac{1}{2l(l+1)}\sum_{\alpha \neq \beta} J_{\alpha \beta} = \frac{F_2+F_4}{14},
	\end{align}
	where $\alpha,\beta$ are orbital indices and $l$ is the angular momentum quantum number.
	
	As mentioned in the Results section, due to a band crossing of the Ru $e_g$ bands with a Na $3s$ band in NaRuO$_2$, there are two sensible ways to select the relevant orbitals considered in the Wannier projection. We denote results based on the five Ru $4d$ orbitals as $(U_{\rm avg},J_{\rm avg})_{4d}$ and results including also the Na $3s$ band as $(U_{\rm avg},J_{\rm avg})_{4d+3s}$. Since both these options lead to very similar results, further calculations in the main text adopt the average of both cRPA results.

	\vspace{0.2cm}{\bf DFT-based derivation of the magnetic model}
	
	\vspace{0.2cm}
	
	We extract the dominant magnetic Heisenberg couplings via the total energy mapping analysis (TEMA)~\cite{totalenergy_analysis0,total_analysis,total_analysis1}, which is a two step process.
	First, we calculate total energies within DFT (GGA+U) of different magnetic configurations of chosen supercells of NaRuO$_2$.
	In the second step we fit the DFT energy of the different magnetic configurations to an effective Heisenberg spin-$1/2$ Hamiltonian using the method of least squares.
	The first step is performed in the VASP 5.3 framework~\cite{VASP} using spin-polarized DFT+U, where we apply the Dudarev scheme~\cite{dudarev}, with effective Coulomb repulsion \text{$U_{\rm eff}$} = $3.5$\,eV.
	Here, we consider $14$ different magnetic configurations. 
	The calculations are performed within a 3$\times$2$\times$1 super cell on 5$\times$8$\times$3 $\Gamma$-centered k-grid with energy cut-off of 540\,eV for the plane-wave basis set.
	The quality of the total energy mapping analysis for the considered model is discussed in Supplementary Note~5. We have checked that different values of the effective Coulomb repulsion don't significantly affect the ratio between the various exchange couplings.
	
	As a second method we employ the so-called projED technique~\cite{riedl2019abinitio}.
	The approach consists of two main steps.
	First, an effective $4d$ electronic Hamiltonian $\mathcal{H}_{\rm tot} = \mathcal{H}_{\rm hop}+ \mathcal{H}_{\rm U}$ is constructed, where $\mathcal{H}_{\rm hop}$ consists of complex electronic hopping parameters, determined from first principles via Wannier projection of a relativistic band structure calculation (GGA+SOC).
	Here, we extract the Wannier functions by using the full potential local orbital (FPLO)~\cite{fplo} package 18.00-57. $\mathcal{H}_{\rm U}$ contains the electronic two-particle Coulomb interaction~\cite{riedl2019abinitio}. 
	In a second step, the electronic Hamiltonian is solved by exact diagonalization on a two-site five-orbital cluster and its low energy states are projected onto spin operators, arriving at the desired effective spin Hamiltonian, e.g. $\mathcal{H}_{2} = \mathbb{P} \mathcal{H}_{\rm tot}\mathbb{P} = \sum_{i<j} \sum_{\mu\nu} \mathbb{J}_{i,j}^{\mu\nu} {S}_i^\mu   {S}_j^\nu$.
	Note that here $\mathbf{S}$ is a pseudospin with $j_{\rm eff}=1/2$. We employ the projED method for the calculation of nearest-neighbor couplings, while for longer-range interactions we resort to TEMA results. This choice is motivated by the fact that, within projED, the indirect hoppings over multiple sites, which are expected to become more and more important for longer-range couplings, cannot be accounted for due to computational limitations. 
	
	We also employ the projED method to extract the four-spin ring exchange Hamiltonian $\mathcal{H}_4^\mathrm{tot}$. Due to computational limitations, the parameters are extracted by diagonalizing a four-site three-orbital electronic Hamiltonian involving only Ru $t_{2g}$ orbitals. We adopt this approximation since the aim of this work is to estimate the general form and order of magnitude of the ring exchange interaction in a strongly spin-orbit coupled system like NaRuO$_2$. Possible refinements of this approach are beyond the scope of this work and will be pursued in future studies.

	\vspace{0.2cm}{\bf Iterative minimization (classical spins)} 
	
	\vspace{0.2cm}
	
	We obtain the classical ground state of the spin Hamiltonian $\mathcal{H}$ by performing a numerical minimization of the energy on a finite lattice with periodic boundary conditions. We employ an \textit{iterative method} in which the orientation of the spins (unit vectors at the classical level) is initialized with random values and updated by performing local moves. A single update is performed by selecting a random site $i$ and changing its spin orientation according to
	\begin{equation}
		\mathbf{S}_i \mapsto -\frac{\mathbf{h}_i}{ \|
			\mathbf{h}_i \| }  \ \mbox{ where } \ \mathbf{h}_i=\left(\frac{\partial\mathcal{H}}{\partial S_i^x},\frac{\partial\mathcal{H}}{\partial S_i^y},\frac{\partial\mathcal{H}}{\partial S_i^z}\right).
	\end{equation}
	In other words, we anti-align the spin at site $i$ to the effective field  $\mathbf{h}_i$ created by the interactions with the other spins in the lattice. The procedure is repeated several times until the minimum of the energy is reached. To try mitigating the possibility of ending up in local energy minima, we perform a number of different calculations starting from different random initializations. Most numerical results have been obtained on a triangular lattice of $N=12 \times 12=144$ sites. For calculations involving inter-layer couplings we used a three-dimensional cluster of $N=6\times 6 \times 6 =216$ sites.

	\vspace{0.2cm}
	{\bf Exact diagonalization}
	
	\vspace{0.2cm}
	
	We perform exact diagonalization of the $j_\text{eff}=1/2$ model on two-dimensional clusters of up to $N=27$ sites. 
	The inelastic neutron scattering intensity at momentum $\mathbf{k}$ and energy $\omega$ is given by 
	\begin{equation}\label{eq:ikomega}
		\small{
			\mathcal{I}(\mathbf{k}, \omega) 
			\propto f^{2}(k) \int \mathrm d t\, e^{-i \omega t} \sum_{\mu, \nu}\left(\delta_{\mu, \nu}-\frac{k_{\mu} k_{\nu}}{k^{2}}\right) \left\langle S_{-\mathbf k}^{\mu}(t) S_{\mathbf k}^{\nu}(0)\right\rangle
		}
	\end{equation}
	where $f(k)$ is the atomic form factor of Ru$^{3+}$. To compute it, we employ the continued fraction method and show results with a Gaussian pole broadening of $\sigma=1$\,meV. To access a higher number of k-points, we plot together results coming from clusters of different shapes and sizes up to $N=27$ (clusters shown in Supplementary Note~3), similar to as done in e.g.\ Ref.~\cite{winter2017breakdown}. 
	
	\vspace{0.2cm}{\bf Linear spin-wave theory}
	
	\vspace{0.2cm}
	
	Linear spin-wave theory calculations are performed with the SpinW 3.0 library~\cite{spinwref}. The inelastic neutron scattering intensity is computed by taking the powder-average of Eq.~\eqref{eq:ikomega}.

	\section{Acknowledgements}
	
	We thank V.\ Krewald and I. I.\ Mazin for valuable advice regarding the \textit{ab initio} calculations.  We thank also D. Ceresoli, M.\ Imada, J.\ Kuneš and P.A. Maksimov for fruitful comments and discussions.
	R.V., A.R., K.R., D.A.S.K.\ and F.F.\ gratefully acknowledge support by the Deutsche Forschungsgemeinschaft (DFG, German Research Foundation) for funding through 
	Project No. 411289067 (VA117/15-1), TRR 288 — 422213477 (project A05) and  CRC 1487 - 443703006 (project A01). L.B.\ was supported by the DOE, Office of Science, Basic Energy Sciences under Award No.\ DE-FG02-08ER46524. S.D.W. acknowledges support by DOE, Office
	of Science, Basic Energy Sciences under Award No.\ DE-SC0017752.

	\section{Author Contributions}
	
	R.V., S.D.W. and L.B. conceived the project. Density functional theory calculations were performed by A.R., K.R., cRPA calculations by S.B., projED calculations by K.R. and calculations on magnetic models by D.A.S.K. and F.F. All authors contributed to the writing of the manuscript.

\bibliography{ref}
	
\clearpage



\section*{Supplementary material}

# A j_eff=1/2 Kitaev material on the triangular lattice: The case of NaRuO₂

Aleksandar Razpopov,[1] David A. S. Kaib,[1] Steffen Backes,[2,3,4] Leon Balents,[5]
Stephen D. Wilson,[6] Francesco Ferrari,[1] Kira Riedl,[1] and Roser Valentí[1]

[1]*Institut für Theoretische Physik, Goethe-Universität, 60438 Frankfurt am Main, Germany*
[2]*Research Center for Advanced Science and Technology,*
*University of Tokyo, Komaba, Tokyo 153-8904, Japan*
[3]*Center for Emergent Matter Science, RIKEN, Wako, Saitama 351-0198, Japan*
[4]*CPHT, CNRS, École polytechnique, Institut Polytechnique de Paris, 91128 Palaiseau, France*
[5]*Kavli Institute for Theoretical Physics, University of California, Santa Barbara, California 93106, USA*
[6]*Materials Department, University of California, Santa Barbara, California 93106-5050, USA*
(Dated: January 18, 2023)

## Supplementary Note 1: Relation of $j_{\text{eff}}$ and $j$ states

In general, the relativistic basis of $d$ block electrons consists of $j = \{5/2, 3/2\}$ states. However, in an octahedral environment, the $e_g = \{x^2\text{-}y^2, z^2\}$ orbitals may be high enough in energy, to make a description in terms of $j_{\text{eff}} = \{3/2, 1/2\}$ states more appropriate [1, 2]. Here, we revisit briefly the relation between these two representations. The $j = 3/2$ states consist of a sum of $j_{\text{eff}} = 3/2$ states and $e_g$ states, as evident from the following expressions:

$$\left|\frac{3}{2}, +\frac{3}{2}\right\rangle = +\sqrt{\frac{3}{5}}i\left|\frac{3}{2}, -\frac{1}{2}\right\rangle_{\text{eff}} + \sqrt{\frac{2}{5}}\left|x^2\text{-}y^2, \downarrow\right\rangle \quad (1)$$

$$\left|\frac{3}{2}, +\frac{1}{2}\right\rangle = -\sqrt{\frac{3}{5}}i\left|\frac{3}{2}, -\frac{3}{2}\right\rangle_{\text{eff}} - \sqrt{\frac{2}{5}}\left|z^2, \uparrow\right\rangle \quad (2)$$

$$\left|\frac{3}{2}, -\frac{1}{2}\right\rangle = -\sqrt{\frac{3}{5}}i\left|\frac{3}{2}, +\frac{3}{2}\right\rangle_{\text{eff}} + \sqrt{\frac{2}{5}}\left|z^2, \downarrow\right\rangle \quad (3)$$

$$\left|\frac{3}{2}, -\frac{3}{2}\right\rangle = +\sqrt{\frac{3}{5}}i\left|\frac{3}{2}, +\frac{1}{2}\right\rangle_{\text{eff}} - \sqrt{\frac{2}{5}}\left|x^2\text{-}y^2, \uparrow\right\rangle \quad (4)$$

In contrast, the information about $j_{\text{eff}} = 1/2$ states is included in a linear combination of $j = 5/2$ states:

$$\left|\frac{1}{2}, +\frac{1}{2}\right\rangle_{\text{eff}} = \sqrt{\frac{1}{6}}i\left|\frac{5}{2}, +\frac{5}{2}\right\rangle - \sqrt{\frac{5}{6}}i\left|\frac{5}{2}, -\frac{3}{2}\right\rangle \quad (5)$$

$$\left|\frac{1}{2}, -\frac{1}{2}\right\rangle_{\text{eff}} = \sqrt{\frac{1}{6}}i\left|\frac{5}{2}, -\frac{5}{2}\right\rangle - \sqrt{\frac{5}{6}}i\left|\frac{5}{2}, +\frac{3}{2}\right\rangle \quad (6)$$

## Supplementary Note 2: Bilinear Hamiltonian in the crystallographic reference frame

In the main text the bilinear Hamiltonian is given in cubic coordinates (shown in Fig. 1b of the main text). Another possible spin coordinate frame, dubbed *crystallographic* frame, is the one in which spin coordinates $(\tilde{S}^x, \tilde{S}^y, \tilde{S}^z)$ co-align with crystallographic $(a, b^*, c)$ directions, as used e.g. in Ref. [3] (and labelled as "cryst" in the following). In the latter framework, focusing on nearest-neighbor bonds, for which anisotropic interactions are relevant, the Hamiltonian

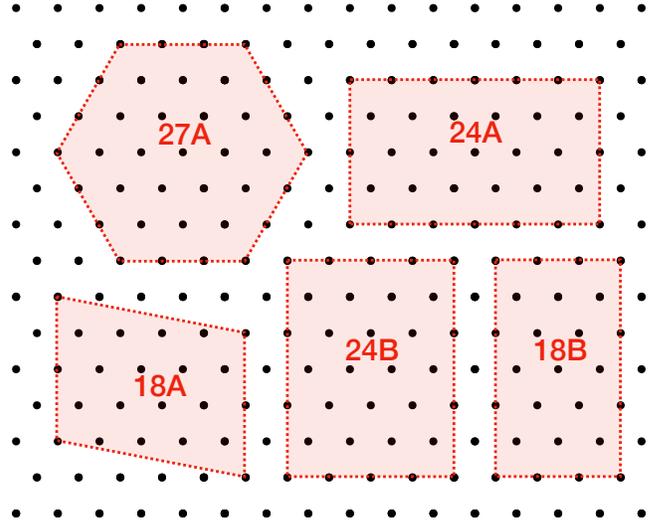

Supplementary Figure 1. **Finite-size clusters used in exact diagonalization calculations.** Dotted lines represent periodic boundaries. Respective number of sites is denoted within each cluster.

reads

$$\begin{aligned}
\mathcal{H}_2^{\text{cryst}} = \sum_{\langle i,j \rangle} \{ &J^{\text{cryst}}(\tilde{S}_i^x \tilde{S}_j^x + \tilde{S}_i^y \tilde{S}_j^y + \Delta^{\text{cryst}} \tilde{S}_i^z \tilde{S}_j^z) \\
&+ 2J_{\pm,\pm}^{\text{cryst}}[(\tilde{S}_i^x \tilde{S}_j^x - \tilde{S}_i^y \tilde{S}_j^y)c_\alpha - (\tilde{S}_i^x \tilde{S}_j^y + \tilde{S}_i^y \tilde{S}_j^x)s_\alpha] \\
&+ J_{z,\pm}^{\text{cryst}}[(\tilde{S}_i^y \tilde{S}_j^z + \tilde{S}_i^z \tilde{S}_j^y)c_\alpha - (\tilde{S}_i^x \tilde{S}_j^z + \tilde{S}_i^z \tilde{S}_j^x)s_\alpha] \},
\end{aligned}$$

(7)

where $c_\alpha = \cos(\varphi_\alpha)$, $s_\alpha = \sin(\varphi_\alpha)$ and $\varphi_\alpha = (0, \frac{2\pi}{3}, -\frac{2\pi}{3})$ for X₁-, Y₁- and Z₁-bonds, respectively [3]. In the crystallographic reference frame, the nearest-neighbor couplings of NaRuO₂ are $(J^{\text{cryst}}, \Delta^{\text{cryst}}, J_{\pm,\pm}^{\text{cryst}}, J_{z,\pm}^{\text{cryst}}) = (-5.97, -0.39, -2.59, -1.65)$ meV, where we note the unusual negative sign of the $XXZ$ anisotropy parameter $\Delta^{\text{cryst}}$.

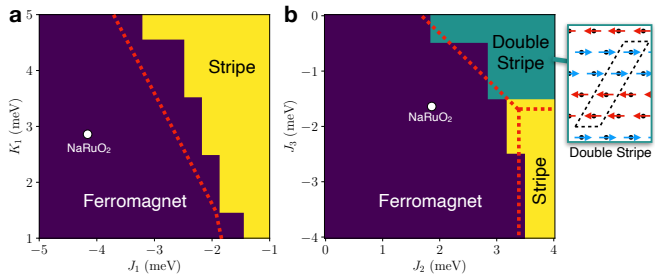

**Supplementary Figure 2.** **Phases of the magnetic bilinear Hamiltonian around the ab-initio estimate for NaRuO₂** Phase diagram of the (in-plane) bilinear Hamiltonian $\mathcal{H}_2$ when tuning **a** the nearest-neighbors interactions $J_1$ and $K_1$, **b** the longer-range couplings $J_2$ and $J_3$. The remaining parameters are fixed and equal to the ones given in Fig. 1d of the main text. The white circles indicate the *ab-initio* derived Hamiltonian for NaRuO₂. The colormap shows the phases obtained by ED calculations (on a coarse finite grid of values, cluster 24A), while the red dashed lines mark the classical phase boundaries. The inset on the right of **b** shows the periodic arrangement of the spins in the double stripe phase (the magnetic unit cell is marked with dashed lines).

## Supplementary Note 3: Details of Exact Diagonalization calculations

In Supplementary Figure 1 we show the various clusters employed in exact diagonalization (ED) calculations. The neutron scattering intensity shown in Fig. 3b of the main text combines results from different momenta allowed on different clusters. For momenta that exist on multiple clusters, the averaged intensity from all such clusters is shown.

## Supplementary Note 4: Perturbations to magnetic Hamiltonian

As discussed in the main text, the magnetic Hamiltonian derived in this study yields a ferromagnetic (FM) ordered ground state. However, different experimental samples might have certain variations in their lattice structure [4], changing their magnetic couplings. Furthermore, the *ab-initio* calculation of the magnetic Hamiltonian depends on input parameters and approximations to some degree. It is therefore interesting to consider how stable the FM ground state of the model is against modest perturbations. We here consider for simplicity only our bilinear model $\mathcal{H}_2$ with intra-plane interactions. This precise model is not captured by existing phase diagrams [3, 5] due to the presence of finite $\Gamma'$, negative $\Delta^{\mathrm{cryst}}$ (see Supplementary Note 2), and finite further-neighbor $J_2$, $J_3$ interactions. We therefore perform various classical energy minimization and ED calculations perturbing around the *ab-initio* derived Hamiltonian. Two phase diagrams are shown in Supplementary Figure 2, in which either $K_1$ and $J_1$, or $J_2$ and $J_3$, are tuned around the NaRuO₂ in-plane bilinear model (marked by the white circle). In all cases, changes of multiple meV to the pa-

rameters in the Hamiltonian are needed to destabilize the ferromagnetic state. This amount of variation exceeds the typical degree of uncertainty expected in the *ab-initio* calculations, supporting the assessment of a pristine NaRuO₂ compound to be ferromagnetic. Interestingly, we note that a "double-stripe" magnetic order is relatively proximate to the NaRuO₂ model. The "double-stripe" order resembles the "stripe" phase, but with a doubled unit cell size, as shown in the inset of Supplementary Figure 2.

We also consider slight variations in the $U$ and $J_H$ (Hund's coupling) parameters away from the cRPA-computed values, which are used in the projED derivation for the magnetic Hamiltonians. However, modest variations of these parameters (e.g., $\pm 0.1$ eV for $J_H$ and $\pm 1$ eV for $U$) still lead to magnetic Hamiltonians with a FM ground state. With decreasing $U$, the (anisotropic) ring exchange interaction tends to increase, however not enough to destabilize the FM order for realistic $U$ and $J_H$ values.

## Supplementary Note 5: Quality of total energy mapping analysis

To quantify the quality of the total energy mapping analysis we consider the difference between the energy of the DFT (GGA+U) calculation and the one provided by the fitting magnetic model. The result is shown in Supplementary Figure 3 for each configuration (labelled by a number). Small deviations between the model calculations and DFT calculations are expected to be accounted for by longer range interactions which are not considered in the model. The small total energy difference (0.049 meV/Ru per configuration) suggests that all relevant exchange couplings are included in the magnetic model.

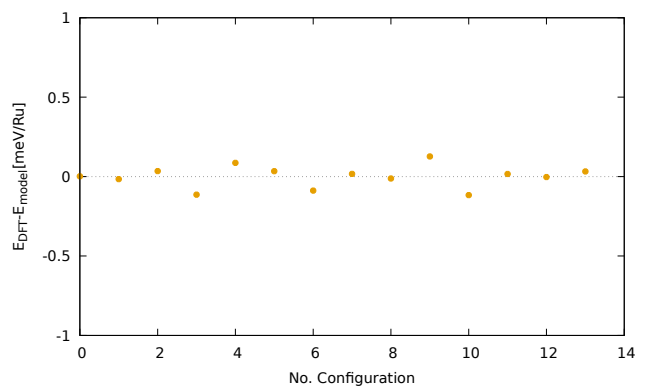

**Supplementary Figure 3.** **Quality of the total energy mapping analysis.** The difference between the total DFT (GGA+U) energy and the one from the magnetic model is shown as a function of the different magnetic configurations, labelled by an integer number.



\end{document}